\documentclass[preprint2]{aastex}
\usepackage{natbib}
\received{August 30, 1999}
\accepted{December 2, 1999}
\slugcomment{Ap. J. in print}

\shorttitle{Energetics of CMEs}
\shortauthors{Vourlidas, Subramanian, Dere}
\begin{document}
\title{LASCO Measurements of the Energetics of Coronal Mass Ejections}
\author{A. Vourlidas, P. Subramanian}
\affil{Center for Earth Observing and Space Research, Institute for
Computational Sciences, George Mason University, Fairfax, VA 22030}
\email{avourlid@pythia.nrl.navy.mil}
\and
\author{ K. P. Dere, R. A. Howard}
\affil{E. O. Hulburt Center for Space Research, Naval Research
Laboratory, Washington, DC~20375} 
\begin{abstract}
We examine the energetics of Coronal Mass Ejections (CMEs) 
with data from the LASCO
coronagraphs on SOHO. The LASCO observations provide fairly direct
measurements of the mass, velocity and dimensions of CMEs. Using these
basic measurements, we determine the potential and kinetic energies
and their evolution for several CMEs that exhibit a flux-rope
morphology. Assuming flux conservation, we use
observations of the magnetic flux in a variety of magnetic clouds near
the Earth to determine the magnetic flux and magnetic energy in CMEs
near the Sun. We find that
the potential and kinetic energies increase at the expense of the
magnetic energy as the CME moves out, keeping the total energy roughly 
constant. This demonstrates that flux rope CMEs are 
magnetically driven. Furthermore, since their total energy is constant,
the flux rope parts of the CMEs can be considered to be a closed system
above $\sim$ 2 $R_{\sun}$.

\end{abstract}
\section{Introduction}

Material ejections are a common phenomenon of the solar
corona. Since the first observation on 14 December 1971
\citep{tousey73}, several thousands of CMEs have been seen
\citep{howard85,kahler92, webb92, hund97, gos97}. Nevertheless, the
mechanisms that cause a CME and the forces acting on it during its
subsequent propagation through the corona are largely unknown.  Of
these two issues, the issue of CME propagation through the corona is
by far more amenable. Past observations have provided
insufficient coverage of the CME development for several reasons:
restricted field of view of the coronagraphs, frequent orbital nights
and low sensitivity of the instruments. Consequently, past studies
were largely focused on either the phenomenological description and
classification of CMEs or the measurement of average values for the
physical properties of the events such as speed, mass, kinetic energy
\citep{jackson78, howard85}. The study of the CME energetics,
in particular, was necessarily restricted to a handful of well observed
events \citep{rust80, webb80}. Their analysis revealed the importance
of the (elusive) magnetic energy and established that 
 the potential energy dominates the kinetic energy. It was also
found that the energy residing in shocks, radio continua and other
forms of radiation was insignificant in comparison to the mechanical
energy of the ejected material. 

The lessons learned from the past resulted in a greatly improved set
of instruments; the LASCO coronagraphs \citep{bru95}, aboard the SOHO
spacecraft \citep{dom95}. The location of the spacecraft at the L1
point permits the continuous monitoring of the Sun while the
combination of the three LASCO coronagraphs provides an unprecedented
field of view from 1.1 $R_\odot$ to 30 $R_\odot$. The replacement of
videcons with CCD detectors and the very low stray light levels of the
coronagraphs have led to a vast sensitivity improvement. It is now
possible to routinely follow the dynamical evolution of a CME.  Here,
we compute basic quantities; mass, velocity and geometry and derive
quantities such as the potential, kinetic and magnetic energies of
CMEs as they progress through the outer corona into the heliosphere.
To our knowledge, this is the first time that detailed observations of the
dynamical evolution of these quantities has been presented. These
measurements are expected to provide concrete observationally-based
constraints on the driving forces in CME models. For this study, we
focus on a group of CMEs that share a common characteristic; namely,
they resemble a helical flux rope in the C2 and C3 coronagraph
images. We choose these events for three reasons: (i) the area of a
CME that corresponds to the flux rope is usually easily identifiable
in the coronagraph images, (ii) their appearance can be related to the
flux rope structures measured in-situ from Earth-orbiting spacecraft,
and (iii) there has been extensive theoretical and observational
interest for this class of CMEs.

Several CMEs observed with the LASCO instrument exhibit a helical
structure like that of a flux rope \citep{chen97,dere99,wood99}. The
theoretical basis for flux rope configurations in solar and
interplanetary plasmas is well established \citep[e.g.,][]{gold63,
gold83, chen93, low96, kumar96, guo96, wu97}. These treatments
envisage the helical flux rope as a magnetic structure that resides in
the lower corona and erupts to form a CME.  There is some debate about
whether the flux rope is formed before the eruption, or whether it is
formed as a consequence of reconnection processes that lead to the
eruption. These arguments are related to those which consider whether
the reconnection occurs above the sheared arcade which presumably
forms the flux rope, or below it \citep{spyro99}.

Neither the physical mechanisms of the initial driving impulse, nor
the conditions in the corona which determine the subsequent
propagation of the flux rope are very well known from
observations. Theoretical models often rely on educated guesses to
model both the initiation of the CME as well as its propagation
through the corona. Statements about the energetics, or driving forces
behind CMEs are made on these bases; for instance, \citet{chen96} and
\citet{wu97_2} show plots of the variation of kinetic,
potential and magnetic energies of CMEs as calculated from their
models. The measurements we present in this paper are expected to
yield some clues about the validity of the assumptions made in these
models. It may be emphasized that our measurements are made only in
the outer corona (2.5 $R_{\odot}$ - 30 $R_{\odot}$).  They are
therefore not expected to shed much light on the energetics of the
flux rope CMEs immediately following initiation, or on the initiation
process itself.

Our estimates of the magnetic energy of flux-rope CMEs are made on the
basis of in-situ measurements of magnetic clouds near the earth.  This
is because flux-rope CMEs ejected from the Sun are often expected to
evolve into magnetic clouds \citep{rust94, kumar96, chen97,
gopal98}. Conversely, in-situ measurements of magnetic clouds near the
earth suggest that their magnetic field configuration resembles a flux
rope \citep{burlaga88, lep90, farrugia95, marubashi97}.  Radio
observations of moving Type-IV bursts can also probe the magnetic
field in CMEs \citep{steward85, rust80} but they are so rare that
near-Earth measurements are the most reliable estimates of the
magnetic flux. It should be borne in mind, however, that the precise
relationship between CMEs and magnetic clouds and the manner in which
CMEs evolve into magnetic clouds is not very well understood
\citep{dryer96, gopal98}. The main reason for this situation is the
simple observational fact that while CMEs are best observed off the
solar limb, magnetic clouds are measured near the Earth. This issue
will hopefully be addressed in the near future by the next generation
of space-borne instruments.

The rest of the paper is organized as follows: We describe our methods
of measuring the mass and position of a CME and of calculating the
different forms of energy associated with it in \S~2. \S~3 presents
the results of our measurements. We discuss caveats that accompany
these results in \S~4 and draw conclusions in \S~5.

\section{Data Analysis}
\begin{figure*}[!ht]
\begin{center}
\includegraphics[width=16cm]{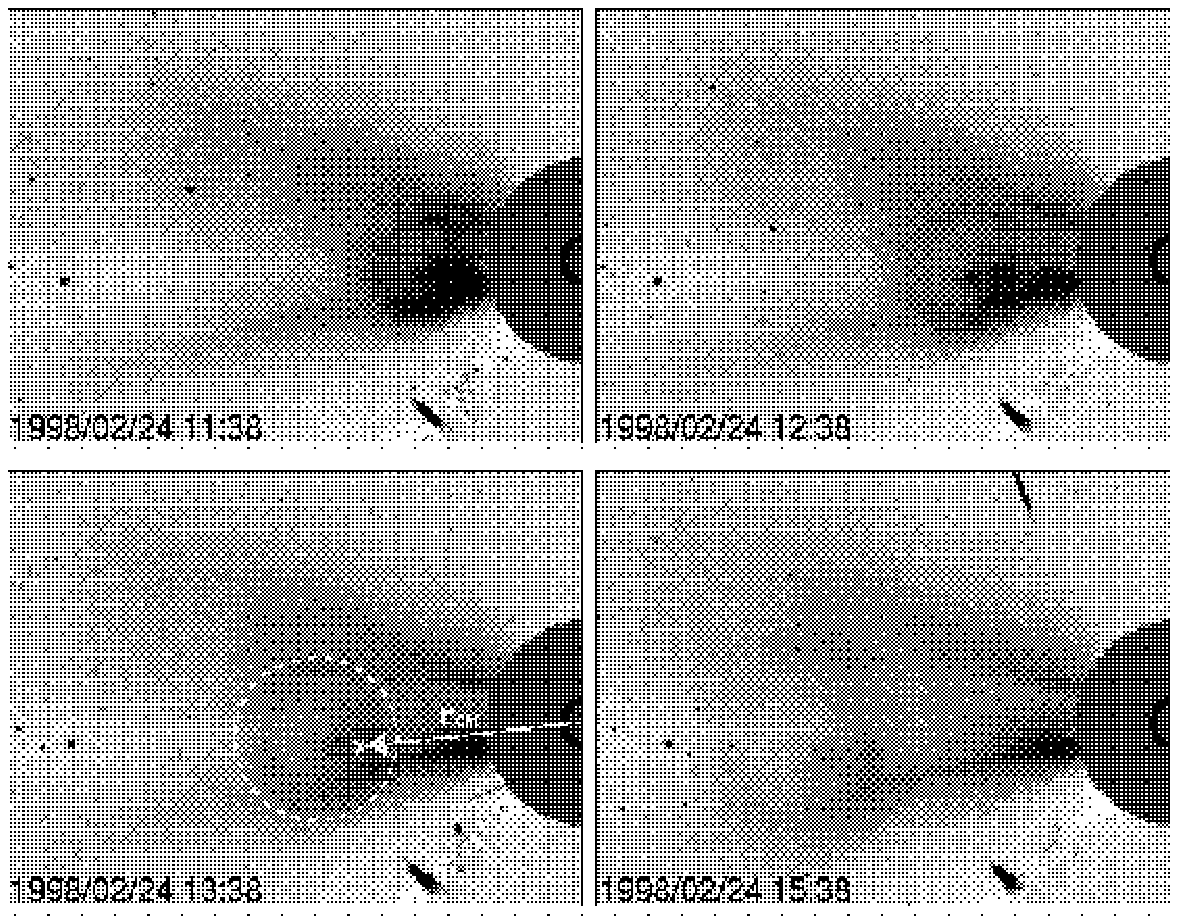}
\caption{ A typical flux-rope CME seen at the eastern half
of the LASCO/C3 coronagraph. The small black circle inside the
occulter represents the solar disk. The helical structure reminiscent
of a flux rope projected on the plane of the sky is easily
discernible.  The dashed circle in the lower left panel demarcates the
flux rope part of the CME used in our calculations. The location of
the center of mass is marked with an 'x' and its vector is
$r_{CM}$.\label{cartoon}}
\end{center}
\end{figure*}
\begin{figure}[!hb]
\includegraphics[width=6cm,angle=90]{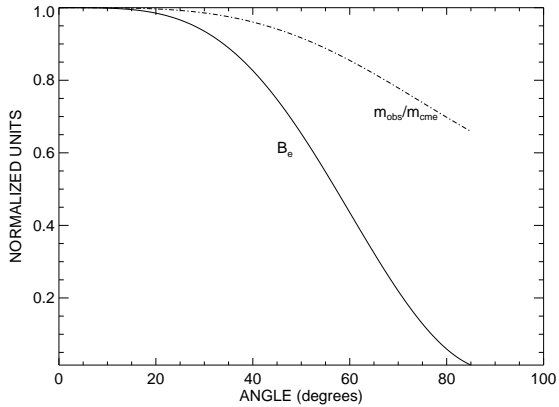}
\caption{ \textit{Solid line:\/} Thomson scattering
calculation of the angular dependence of the total brightness of a
single electron at a heliocentric distance of 10 R$_\odot$. The curve
is normalized to the brightness at $0^\circ$. \textit{Dash-dot
line:\/} Ratio of the observed relative to the actual mass of a
simulated CME, centered on the plane of the sky, as a function of its
angular width (see text).  \label{siml}}
\end{figure}

\subsection{Mass calculations}

White light coronagraphs detect the photospheric light scattered by
the coronal electrons and therefore provide a means to measure coronal
density. Transient phenomena, such as CMEs, appear as intensity
(hence, density) enhancements in a sequence of coronagraph images. We
compute the mass for a CME in a manner similar to that described by
\citet{poland81}. After the coronagraph images are calibrated in units
of solar brightness, a suitable pre-event image is subtracted from the
frames containing the CME. The excess number of electrons is simply
the ratio of the excess observed brightness, $B_{obs}$, over the
brightness, $B_e(\theta)$, of a single electron at some
angle, $\theta$, (usually assumed to be 0) from the plane of the sky. 
$B_e(\theta)$ is computed from the Thomson scattering function
\citep{bill66}. The mass, $m$, is then calculated assuming that the ejected
material comprises a mix of completely ionized hydrogen and 10\%
helium. Namely,
\begin{equation}
m = {B_{obs}\over B_e(\theta)}\cdot 1.97\times10^{-24} {\rm gr}
\end{equation}
 
\noindent After the mass image is obtained, we delineate the flux rope
by visual inspection, as shown in Figure~\ref{cartoon}. We attempt to
circumscribe the cross section of the helical flux rope as seen in the
plane of the sky. The cavity seen in the white light/mass images is
taken to be the interior of the flux rope, bounded by the helical
magnetic field (Figure~\ref{cartoon}).  The mass contained in the flux
rope is computed by summing the masses in the pixels encompassed by
the flux rope.

The accuracy of the mass calculations depends on three factors: the
CME depth and density distribution along the line of sight and the
angular distance of the CME from the plane of the sky. All three
factors are unknown since the white light observations represent only
the projection of the CME on the plane of the sky. Some additional
information can be obtained from pB measurements, but these are only
occasionally available. Therefore, to convert the observed brightness
to a mass measurement we have to make an assumption. Namely, we assume
that all the mass in the CME is concentrated in the plane of the
sky. Since CMEs are three-dimensional structures, our calculations
will tend to underestimate the actual mass. 

To quantify the errors arising from our assumption, we performed two
brightness calculations shown in Figure~\ref{siml}. The solid line
shows the angular dependence of the quantity $B_e(\theta)$ in
equation~(1) normalized to its value at $0^\circ$. We see that our
assumption that the ejected mass is always in the sky plane
($\theta=0^\circ$) underestimates the mass by about a factor of 2 at
angles $\sim 50-60^\circ$. We expect that the CMEs in our sample are
relatively close to the plane of the sky ($\theta<50^\circ$) since
their flux rope morphology is clearly visible.

Next, we investigate the effect of the finite width of a CME. We
simulate a CME with constant density per angular bin along the line of
sight, centered in the plane of the sky at a heliocentric distance of
10 R$_{\sun}$. Using equation (1) we calculate the observed mass,
$m_{obs}$, for various widths and compare it to the actual mass,
$m_{cme}$ for the same widths. The dashed line in Figure~\ref{siml}
shows the dependence of this ratio, $m_{obs}/m_{cme}$ on the width of
the CME. For angular widths similar to those of the CMEs in our sample
($\lesssim60^\circ$) the mass would be underestimated by about
$\sim15\%$.

Finally, we estimate the noise in the LASCO mass images from
histograms of empty sky regions. The statistics in these areas show a
gaussian distribution centered at zero, as expected. We define the noise
level as one standard deviation or about $5\times10^8$ gr in the C2
frames and $3\times10^{10}$ gr in the C3 frames. The average C2 pixel
signal in the measured CMEs is 10 times the noise and the C2 pixel
signal-to-noise ratio in the mass measurements is between 10-100. The
CMEs get fainter as they propagate farther from the sun. Therefore,
the pixel signal-to-noise ratio in the C3 images drops to about
3-4. These figures refer to single pixel statistics and demonstrate
the quality of the LASCO coronagraphs. Our measurements are based on
statistics of hundreds or thousands of pixels for each
image. Therefore, the ``mass'' noise in our images is insignificant
compared to the systematic errors involved in the calculation of a CME
mass as discussed previously.

In summary, these calculations suggest that the LASCO measurements
tend to underestimate the CME mass by about 50\%, for realistic
widths and propagation angles. A more detailed analysis of CME mass
calculations will appear elsewhere.

\subsection{CME Energy calculations}

In this analysis we consider only three forms of energy --- potential,
kinetic, and magnetic energy. These energies can be estimated from
quantities measured directly in the LASCO images like CME area, mass
and speed. Two of the many other forms of energy that can exist in the
CME/corona system can be estimated based on some assumptions and
educated guesses: the CME enthalpy $U$ and the thermal energy
$E_T$. We will show in \S~4 that the thermal energy $E_T$ is
insignificant. There are several uncertainties involved in
calculating the enthalpy of a CME. Firstly, the temperature structure
of a CME is far from known. It is conceivable that is composed of
multithermal material.  In situ measurements of magnetic clouds near
the earth reveal a temperature range of $10^4-10^5$ K. Furthermore, it
is not clear if the gas in the CMEs in the outer corona is in local
thermodynamic equilibrium. Nonetheless, if we assume the CME to be a
perfect gas in local thermodynamic equilibrium with equal electron and
ion temperatures, the enthalpy $U$ can be as large as $5E_T\,=
5nkT$. If we assume a temperature of a million degrees K and a mass of
$10^{15}$ gr, this yields $U\approx 3\times10^{29}$ ergs. As will be
seen later, even this upper limit for the enthalpy $U$ is lower than
the kinetic and potential energies by at least one order of magnitude,
except in the lower corona where it can be comparable to the kinetic
energy. Furthermore, the enthalpy is directly proportional to the
mass, which, as will be seen later, remains approximately constant as
the CME propagates outwards. We therefore conclude that the enthalpy
is a small, constant magnitude correction which can be safely
neglected without affecting the overall conclusions regarding CME
energetics.

\paragraph{Potential Energy}

We define the potential energy of the flux rope as the amount of
energy required to lift its mass from the solar surface. The
gravitational potential energy is calculated using

\begin{equation}
E_P = { \sum_{{\rm flux}\,{\rm rope}}} \,
\int^{R}_{R_{\odot}} \,
\frac{G\, M_{\odot} \, m_{i}}{r_{i}^{2}} \, dr_{i} \, ,
\end{equation}

\noindent where $m_{i}$ and $r_{i}$ denote the mass and distance from
sun-center respectively, of each pixel, $M_{\odot}$ is the mass of the
sun, $R_{\odot}$ is the solar radius and $G$ is the gravitational
constant. The summation is taken over the pixels comprising the flux
rope (Figure~\ref{cartoon}).

\paragraph{Kinetic Energy}

We use the center of mass of the flux rope to describe its
movement. The location of the center of mass relative to the sun
center is given by

\begin{equation}
\vec{r}_{CM} = \frac{{\sum_{{\rm flux}\,{\rm
rope}}}\,m_{i}\,\vec{r_{i}}}{{\sum_{{\rm flux}\,{\rm rope}}}\,m_{i}} \, ,
\end{equation}

\noindent where $\vec{r}_{CM}$ is the radius vector of the center of
mass and $\vec{r_{i}}$ is the radius vector for each pixel. The
summation, as before, is taken over the pixels comprising the flux
rope. We calculate $\vec{r}_{CM}$ for each CME frame as it progresses
through the LASCO field of view. In other words, we compile a table of
center-of-mass locations versus time, ($\vec{r}_{CM}\, , t$). By
fitting a second degree polynomial to ($\vec{r}_{CM}\, , t$) we obtain
the center of mass velocity, $\vec{v}_{CM}$ and acceleration
$\vec{a}_{CM}$.  The calculation of
the speed and acceleration as described above has the advantage of
involving only the measurement of the CME center of mass. Once the
flux rope is delineated, its mass, speed and energetics follow. The
kinetic energy is simply
\begin{equation}
E_K = \frac{1}{2}\, \sum_{{\rm flux}\,{\rm rope}} \, 
m_{i} \, v_{CM}^{2} \, .
\end{equation}
Note that these measurements are based on the plane of the sky
location of the center of mass. The speed used in the calculations is
therefore a projected quantity and not the true radial speed.  It
follows that the derived kinetic energies are lower limits. The same
applies for all of our observed and derived quantities which
facilitates the comparison among the different events.
\paragraph{Magnetic Energy}

The calculations of the potential and kinetic energies of flux rope
CMEs are made directly from the mass images. On the other hand, the
values we use for the magnetic energy of these CMEs are only estimates
because the magnetic field strength in a CME is unknown. In-situ
measurements by spacecraft like WIND yield the magnetic field
contained in magnetic clouds observed near the earth. As mentioned in
\S~1, helical flux-rope CMEs are thought to evolve into magnetic
clouds similar to those observed at the earth. Therefore, measurements
of the magnetic flux contained in such magnetic clouds are expected to
be fairly representative of that carried by flux rope CMEs. The
magnetic energy carried by a flux rope CME is defined by
\begin{equation}
E_M = 
\frac{1}{8\,\pi} \int_{{\rm flux}\,{\rm rope}} B^{2} dV\, ,
\end{equation}
where $B$ is the magnetic field carried by the flux rope, and the
integration is carried out over the volume of the flux rope.  For a
highly conducting medium such as the heliosphere, the magnetic flux,
$\int B dA$, is frozen into the CME as it evolves to form a magnetic
cloud. The magnetic flux measured in-situ is therefore taken to be the
same as that contained in the CME as it passes through the LASCO field
of view. We use this frozen flux assumption since we feel that it is a
simple, physically motivated one. Another assumption which gives very
similar results is conservation of magnetic helicity \citep{kumar96}.
The volume integral in equation (5) contains another unknown; the
volume occupied by the flux rope. Assuming a cylindrical flux rope
with constant magnetic field, equation (5) is approximated as
\begin{equation}
E_M \sim \frac{1}{8\,\pi}\, \frac{l}{A} \, (B\cdot A)^2  ,
\end{equation}
where $A$ is the area of flux rope as measured in the LASCO images and
$l$ is the length of flux rope. The quantity $B\cdot A$ is the
magnetic flux frozen into the flux rope and is conserved. For our
purposes, we need, in equation (6), a representative value for the
magnetic flux of a flux rope. We obtain such an estimate from model
fits \citep{lep90} to several magnetic clouds observed by WIND between
1995--1998 available at
\url{http://lepmfi.gsfc.nasa.gov/mfi/mag\_cloud\_pub1p.html}. We only
consider clouds that occurred at the same time interval as the LASCO
CMEs (1997-98). From this sample we get the average magnetic flux,
$<B\cdot A> = 1.3\pm1.1 \times10^{21}$ G cm$^2$ which we put in
equation (6). The resulting magnetic energy uncertainty is then
$(1.1/1.3)^2 \approx 70\%$. To calculate the magnetic energy, we also
need the length $l$ of the rope along the line of sight. Since the
true length of the rope cannot be obtained observationally, we assume
that the flux rope is expanding in a self-similar manner, with its
length being proportional to its heliocentric height; namely, $l\sim
r_{CM}$.
\begin{deluxetable}{ccccc}
\tablecaption{CME Event List}
\tablehead{\colhead{Date} & \colhead{First C2 appearance} &
\colhead{Position angle} & \colhead{Angular Width} & \colhead{Final Speed}\\
\colhead{} & \colhead{ (UT)} & \colhead{CCW from North (Deg)} & \colhead{(Deg)}
& \colhead{(km/sec)}
}
\startdata
970223\tablenotemark{a} \tablenotemark{b} & 02:55 & 89  & 63 & 920\\
970413\tablenotemark{c} & 16:12 & 260 & 42 & 520\\
970430\tablenotemark{a} & 04:50 & 83  & 70 & 330\\
970813\tablenotemark{d} & 08:26 & 272 & 36 & 350\\
971019\tablenotemark{b} & 04:42 & 90  & 77 & 263\\
971030 & 18:21\tablenotemark{f} & 85  & 50 & 215\\
971031 & 09:30 & 260 & 54 & 476\\
981101\tablenotemark{b} & 20:11 & 272 & 57 & 264\\
980204 & 17:02 & 284 & 43 & 420\\
980224 & 07:55 & 88  & 32 & 490\\
980602\tablenotemark{e} & 09:37 & 246 & 47 & 600\\
\enddata
\tablenotetext{a}{\citet{wood99}}
\tablenotetext{b}{\citet{dere99}}
\tablenotetext{c}{\citet{chen97}}
\tablenotetext{d}{\citet{andrews99}}
\tablenotetext{e}{\citet{plunk99}}
\tablenotetext{f}{The time refers to the previous day, 97/10/29.}
\end{deluxetable}

Finally, we emphasize that the magnetic cloud data used here are only
representative. They are not measurements from the same LASCO events
we analyzed. Also the magnetic flux in individual events can differ
from the average value we adopted. Furthermore, the magnetic field
values we use refer to the total (toroidal + poloidal) magnetic field
contained in the flux rope. The definition of $B\cdot A$, however,
refers only to the toroidal component of the magnetic field which is
normal to the cross-sectional area of the flux rope. For these
reasons, it is difficult to ascribe errors to our magnetic energy
calculations of individual events. Therefore, we decided to use the
statistical uncertainty in the average flux to compute the error in
the magnetic energy which is about 70\% as shown above. It is
unfortunate that the magnetic energy measurements are so uncertain and
they will continue to be so until direct observations of the coronal
magnetic field become available.

\section{Results}
\begin{figure*}[p]
\includegraphics[width=15cm]{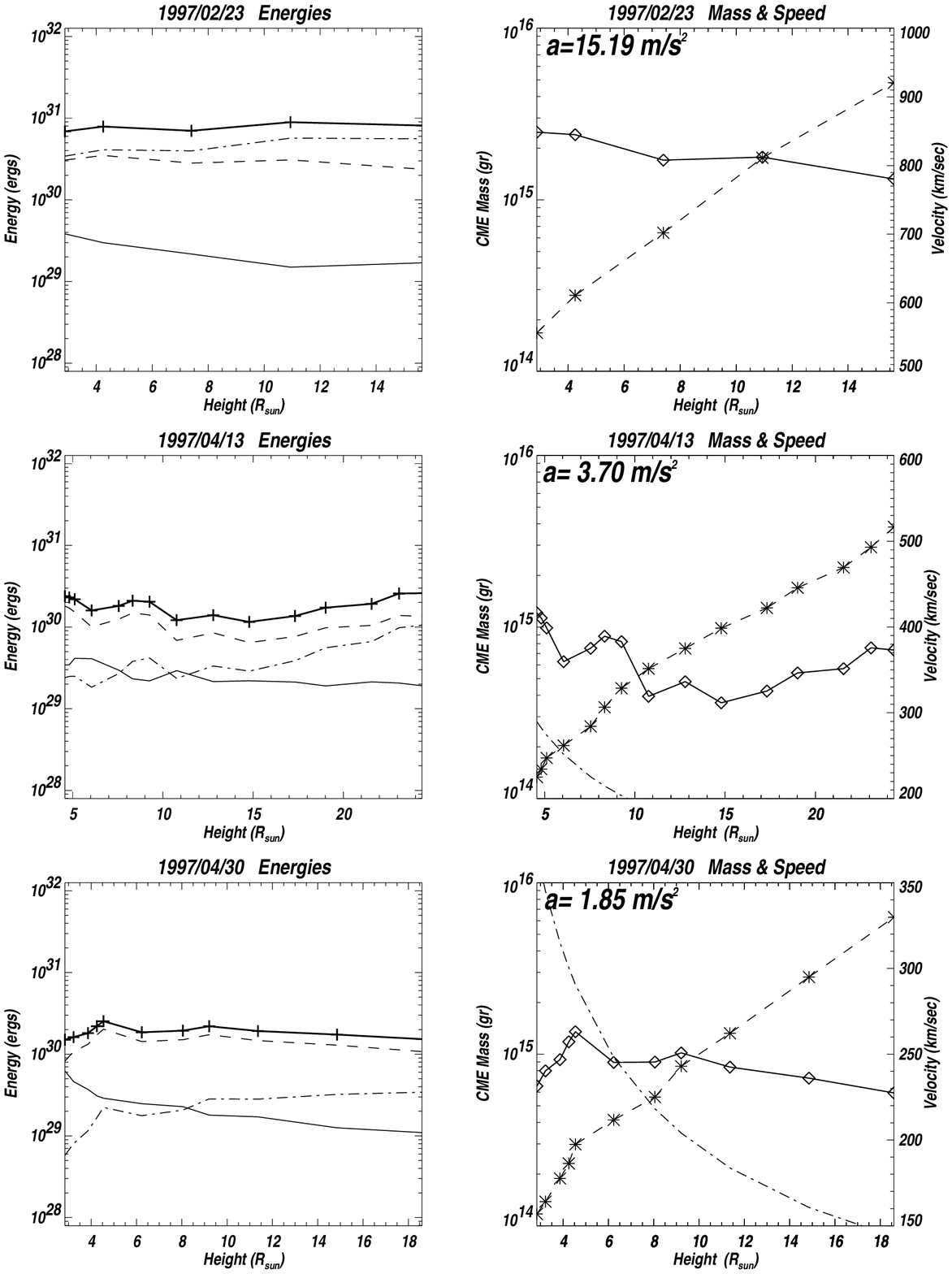}
\caption{ LASCO measurements of CMEs. Each row is a separate
event, labeled by its observation date in the C2 coronagraph. The
horizontal axis denotes heliocentric height (in R$_\odot$).\textsl{
Left panels:\/} Evolution of potential (\textsl{dash}), kinetic
(\textsl{dash-dot}), magnetic (\textsl{solid}) and total
(\textsl{solid with pluses}) CME energies. The total energy is the sum
of the potential, kinetic and magnetic energies.\textsl{ Right
panels:\/} Evolution of the mass (\textsl{solid line with diamonds})
and the center-of-mass speed (\textsl{dashed line with asterisks}).
The derived acceleration is also shown.  The dash-dot line, visible in
some plots, is the escape speed from the Sun as a function of
height.\label{res1}}
\end{figure*}
\begin{figure*}[p]
\includegraphics[width=15cm]{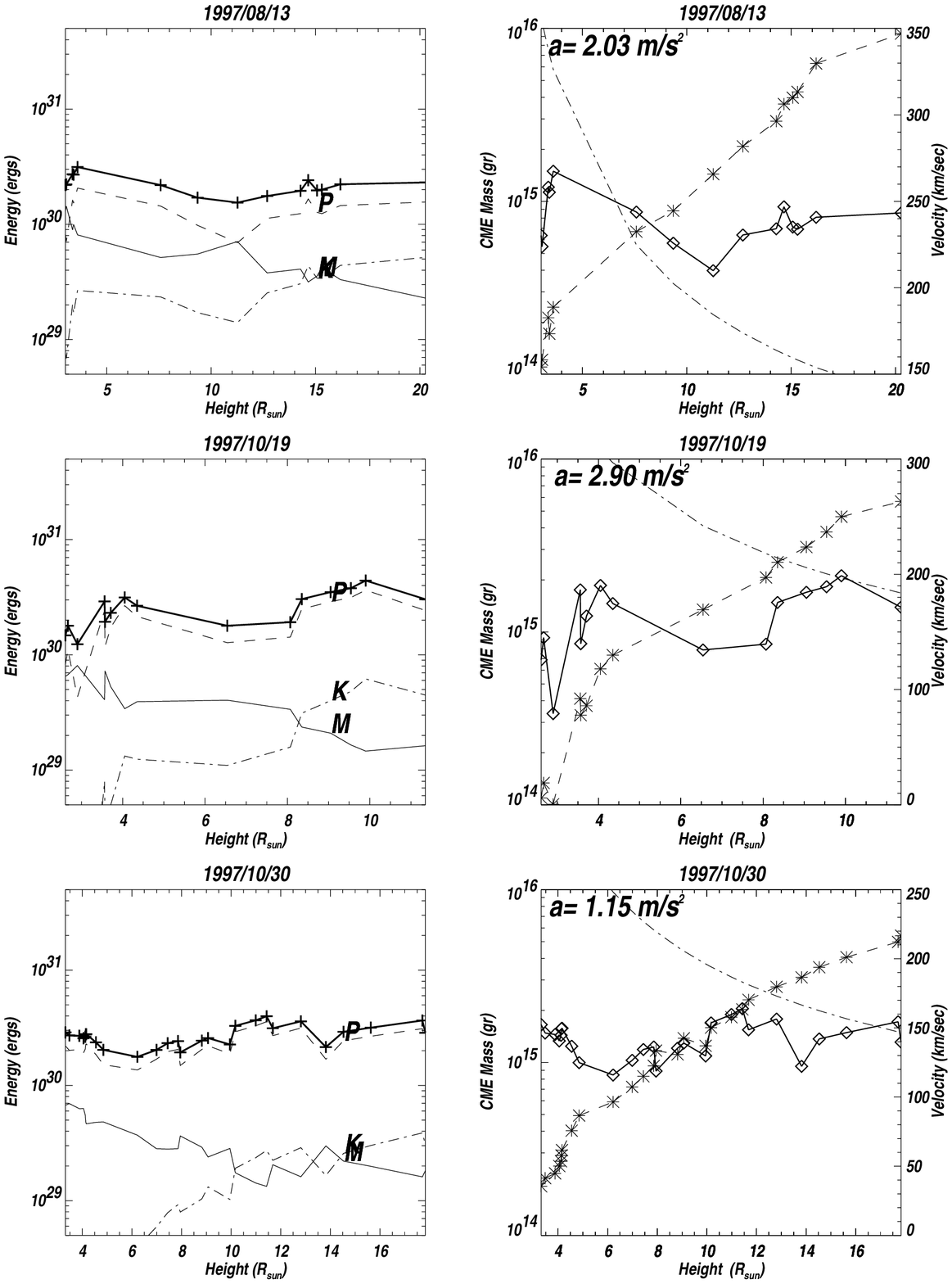}
\caption{Same as Figure~\ref{res1}.\label{res2}}
\end{figure*}
\begin{figure*}[p]
\includegraphics[width=15cm]{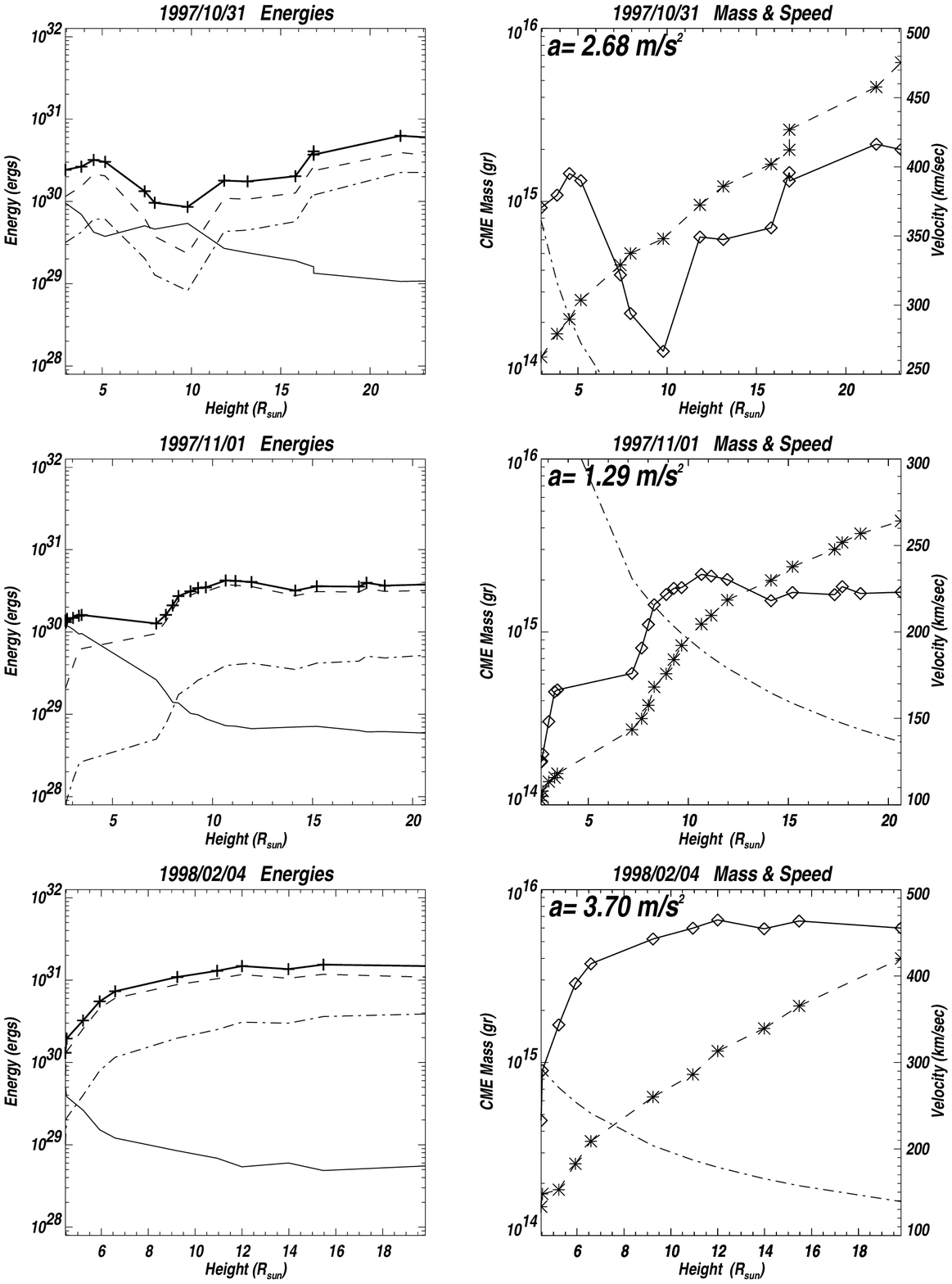}
\caption{Same as Figure~\ref{res1}.\label{res3}}
\end{figure*}
\begin{figure*}[p]
\includegraphics[width=15cm]{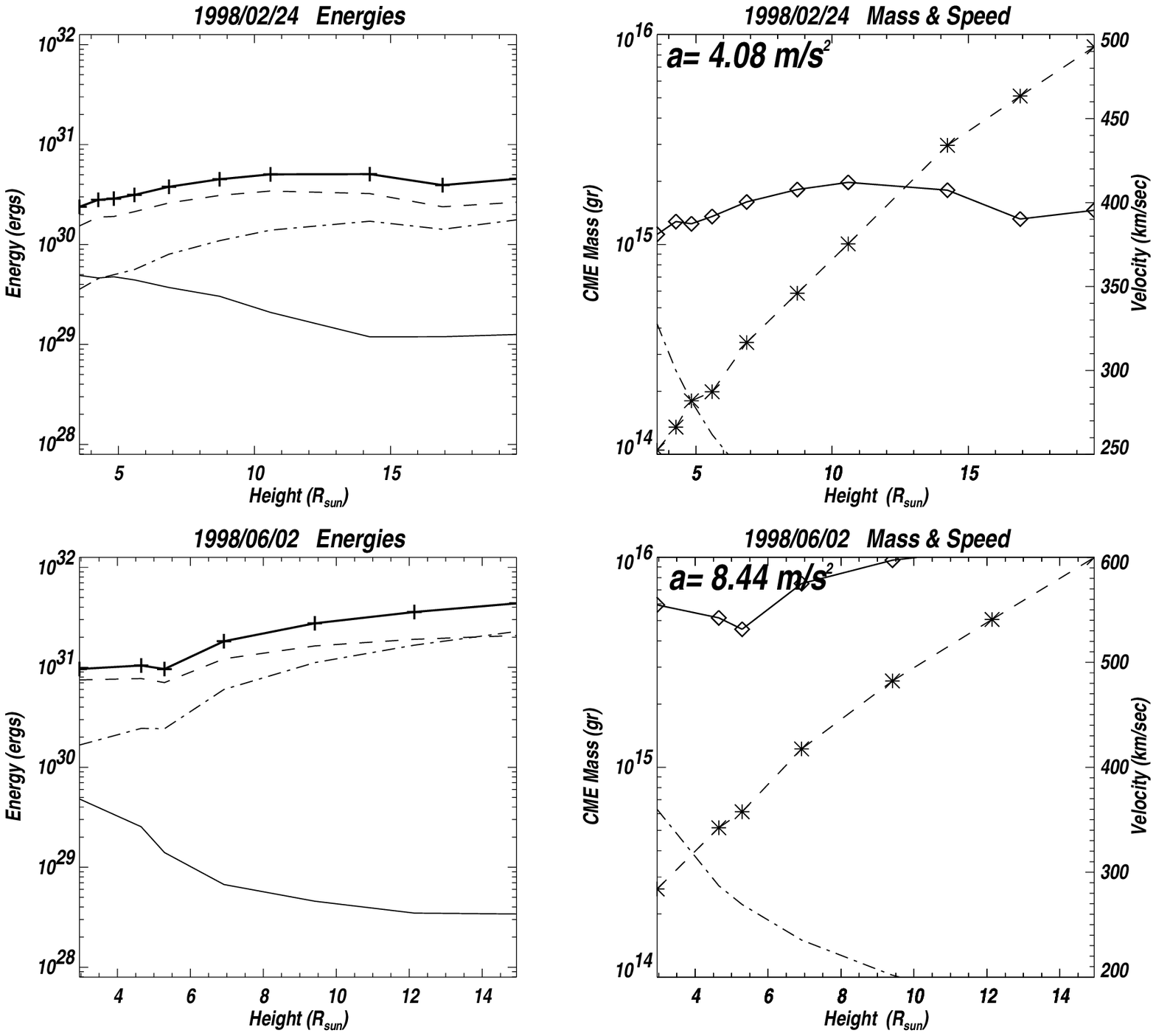}
\caption{Same as Figure~\ref{res1}.\label{res4}}
\end{figure*}

For our analysis, we searched the LASCO database for CMEs with clear
flux rope morphologies. We picked 11 events for which we compiled the
evolution of the mass and velocity of the center of mass and the
potential, kinetic and magnetic energies as the CME progressed through
the LASCO C2 and C3 fields of view. For reference purposes we present
a list of the events in Table~1. The information for the 1997 CMEs is
taken from the LASCO CME list compiled by Chris St.~Cyr
(\url{http://lasco-www.nrl.navy.mil/cmelist.html}) except for the
final speeds in the last column that refer to the center of mass of the
fluxropes and were calculated by us. Further information on source
regions and associated photospheric/low corona emissions for some of
these events can be found in the references noted in the table.

Our measurements are shown in Figures~\ref{res1} -- \ref{res4}. The
horizontal axis denotes heliocentric height in solar radii.  Each row
is a separate CME event, labeled by its date of observation by the
LASCO/C2 coronagraph. The left panels show the evolution of the
potential, kinetic, magnetic and total energy in the CME. The total
energy is the sum of the potential, kinetic and magnetic energies. The
right panels show the evolution of the flux rope mass and the
center-of-mass speed.  As discussed in \S~2, a second degree fit to
($\vec{r}_{CM}\, , t$) yields the acceleration of the center of mass
$\vec{a}_{CM}$.  The radial component of $\vec{a}_{CM}$ is also shown
in this panel.  The dash-dot line, visible in some plots, marks the
escape speed from the Sun as a function of height.

An inspection of the plots leads to the following
overall conclusions that hold for most of the events:

\begin{itemize}

\item
The total energy (curves marked with +) is relatively constant, to
within a factor of 2, for the majority of the events despite the
substantial variation seen in the individual energies.  This suggests
that, for radii between approximately $3 R_{\odot}$ and $30
R_{\odot}$, the flux rope part of these CMEs can be considered as an
isolated system; i.e., there is no additional ``driving energy'' other
than the energies we have already taken into account (potential and
kinetic energies of the flux rope, and magnetic energy associated with
the magnetic field inside the flux rope).

\item
We see that the kinetic and, (to a lesser degree) potential energies
increase at the expense of the magnetic energy, keeping the total energy
fairly constant.  The decrease in magnetic energy is a direct consequence
of the expansion of the CME.  It could imply that the untwisting of the
flux rope might be providing the necessary energy for the outward
propagation of the CME in a steady-state situation.

\item
The center of mass accelerates for most of the events, and the CMEs
achieve escape velocity at heights of around 8-10 $R_{\sun}$, well
within the LASCO/C3 field of view.

\item
The mass in the flux rope remains fairly constant for some events
(e.g., 97/08/13 or 97/10/30) while other events (e.g., 97/11/01 or
98/02/04) exhibit a significant mass increase in lower heights and
tend to a constant value in the outer corona, above about $10-15$
R$_{\sun}$. This observation raises the question: why is pile up of
preexisting material observed only in some flux rope CMEs? We plan to
investigate this effect further in the future. It would also be
interesting to examine how the mass increase close to the Sun relates
to interplanetary ``snowplowing'' observations \citep{webb96}.
\end{itemize}

The only notable exception is the event of 98/06/02 which is also the most
massive and its total energy increases with distance from the center
of the sun. This CME is associated with an exceptionally bright
prominence which may affect the measurements. A detailed analysis of
this event is presented in \citet{plunk99}.

\section{Discussion}

The conclusions of the previous section are based on a set of
broadband white light coronagraph observations. The accuracy of the
measurement of any structure (i.e., CME) in such images is
inherently restricted by three unknowns: the amount and distribution
of the material and the extent of the structure along the line of
sight.

We addressed the first two problems in \S~2 where we showed that for
the case of a uniformly filled CME extending $\pm 80$ degrees
out of the plane of the sky, we will measure about 65\% of its
mass. Since the potential and kinetic energies are directly
proportional to the mass, our measurements in
Figures~\ref{res1} -~\ref{res4} could be higher by as much as 35\%. The
spatial distribution of the material will also affect the visibility
of the structures we are trying to measure. Because we delineate the
area of the flux rope by visual inspection, we might not be
following the same cross section as the structure evolves. This might
account for some of the variability  of the energy curves. However, we
chose the CMEs based on their clear flux rope signatures. The
measurements involve hundreds or even thousands of pixels per
image and therefore we don't expect that the trends seen in the data
are affected by the slight changes in the visibility of the structure.

The widths along the line of sight of the observed CMEs are difficult
to quantify. There is no way to measure this quantity with any
instrumentation in existence today. Only the magnetic energy depends
on the width of the flux rope. In \S~2.2, we assumed that the width
of the flux rope is equal to the height of its center of mass which
implies that its preeruption length is about a solar
radius. Prominences and loop arcades of this length are not uncommon
features on the solar surface. 
\begin{figure*}[!ht]
\includegraphics[width=12cm,angle=90]{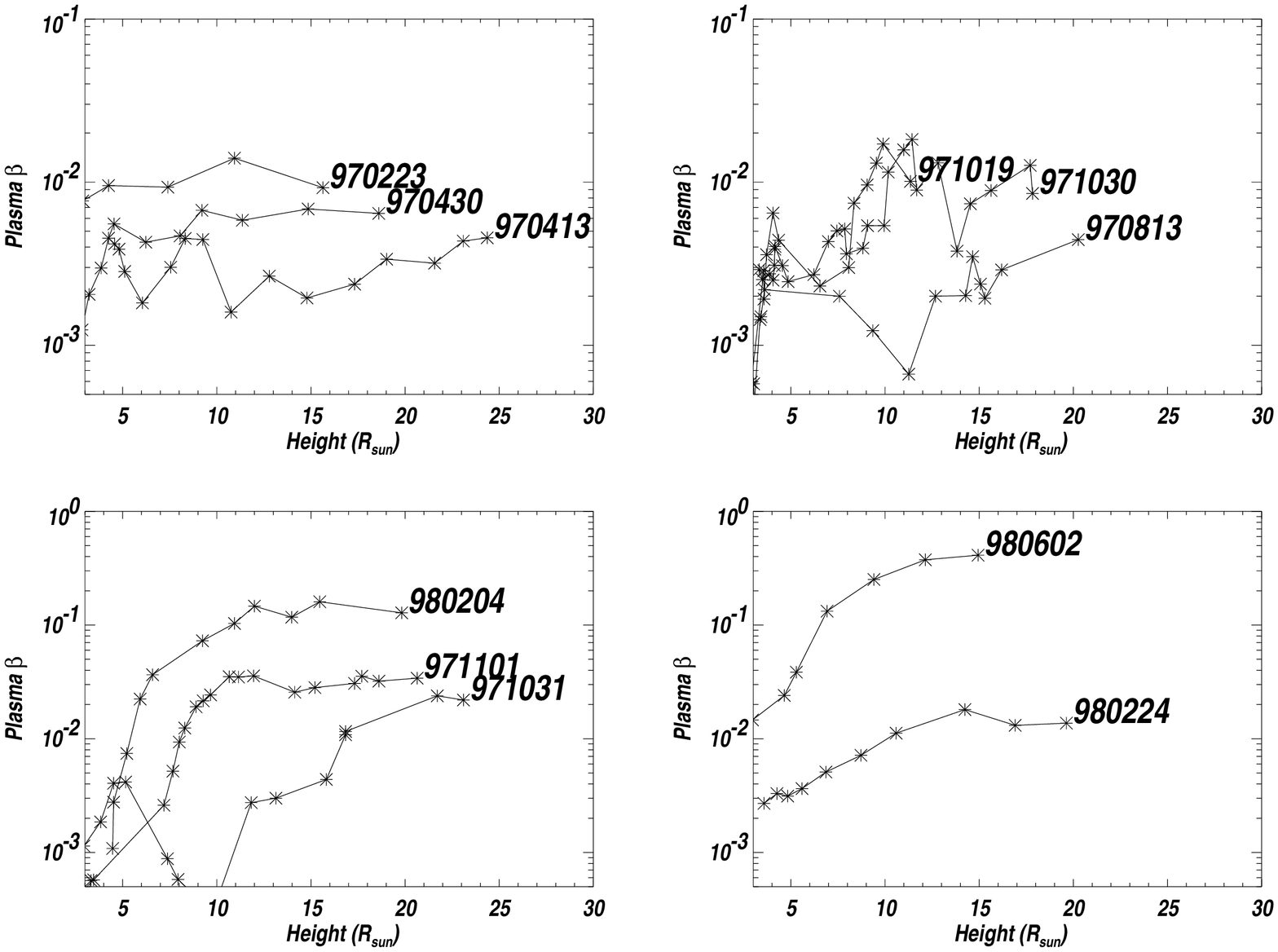}
\caption{Plot of plasma $\beta$ as a function of height
for the flux rope CMEs. We assumed that the CME material is at a
temperature of $10^6$ K.\label{beta}}
\end{figure*}

As described in \S~1, flux rope CMEs are expected to evolve into
magnetic clouds near the earth. This is the basis on which we use
in-situ data to estimate the magnetic energy carried by the flux rope
CMEs (\S~2). In \S~2, we also estimated that the overall normalization
of the magnetic energy curve is uncertain by about 70\%.

In summary, none of the above errors can affect the trends of the
curves for a given event. Only the magnitudes of the various energies
could change. Finally, some of the variability of the measured
quantities could be attributed to the intrinsic variability of the
corona and/or of the CME structure itself and cannot be removed
without affecting the photometry. For this reason, it is rather
difficult to associate an error estimate to individual
measurements. Therefore, we decided not to include any error bars in
our figures.

The analysis of the CME dynamics in Figures~\ref{res1}-\ref{res4}
reveals an interesting trend; namely, the total energy remains
constant. It appears that the flux rope part of a CME propagates as a
self-contained system where the magnetic energy decrease drives the
dynamical evolution of the system. All the necessary energy for the
propagation of the CME must be injected in the erupting structures
during the initial stages of the event. The notions that these CMEs
are indeed magnetically driven and that the thermal energy
contribution can be ignored are further reinforced by the magnitude of
the plasma $\beta$ parameter (Fig.~\ref{beta}). The calculations were
performed with the assumption that the CME material is at a coronal
temperature of $10^6$ K. We see that the CMEs have a very small
$\beta$ (except the events on 98/02/04 and 98/06/02) which increases
slightly outwards. It appears to tend towards a constant value. Such a
behavior for the plasma $\beta$ parameter was predicted in the
flux rope model of \citet{kumar96}. We also find that the potential
energy is larger than the kinetic energy. These results confirm the
conclusions from earlier Skylab measurements (see \citet{rust80} for
details).

The relation between the helical structures seen in the coronagraph
images and eruptive prominences is still unclear. In our sample, only
half of the CMEs have clear associations with eruptive prominences
(e.g., 97/02/23). No helical structures are visible in pre-eruption
EIT 195\AA{} images, in agreement with past work \citep{dere99}. On
the other hand, the flux rope of the event on 98/06/02 is very clearly
located above the erupting prominence and there is strong evidence
that it was formed before the eruption \citep{plunk99}. It seems,
therefore, likely that the process of the formation of the flux rope
is completed during the early stages of the eruption at heights below
the C2 field of view ($<2$ R$_{\sun}$). Such an investigation,
however, is beyond the scope of this paper.
\begin{figure*}[!ht]
\includegraphics[width=12cm,angle=90]{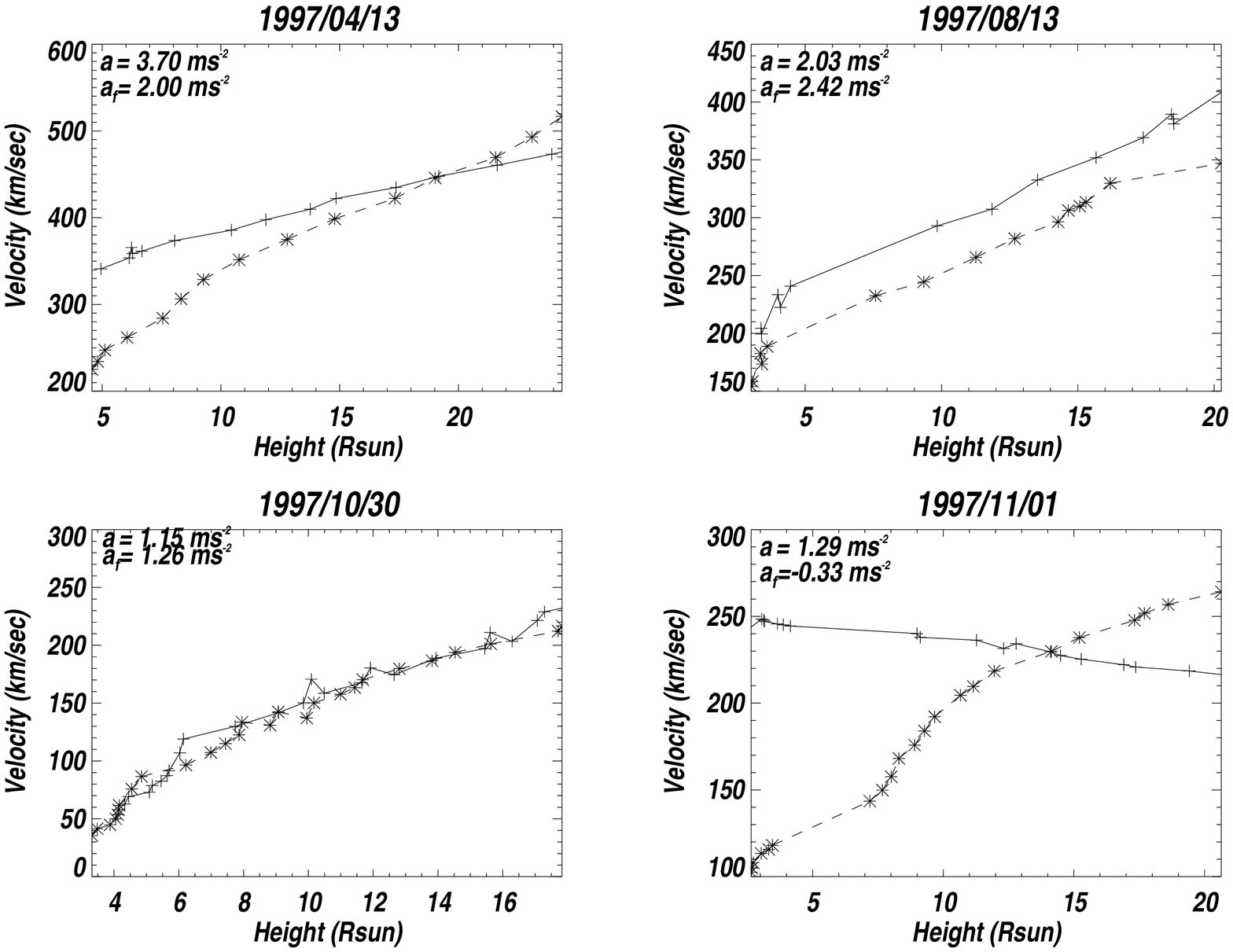}
\caption{Comparison of the speeds between the front
(\textsl{crosses\/}) and the center of mass (\textsl{asterisks\/}) of
the flux rope for four representative events. The accelaration
resulting for a polynomial fit to these curves is also shown where $a$
and $a_f$ are the acceleration of the center of mass and the front,
respectively.\label{front}}
\end{figure*}

Finally, we turn our attention to the evolution of the flux rope shape
as a function of height. We proceed by comparing the velocity of the
CME front to its center of mass velocity. Because the visual
identification of points along the front can be influenced by
visibility changes as the CME evolves, it is susceptible to error. A
better method is to use a statistical measure for the location of the
front such as the center of mass. Hence, the location of the front is
defined as the center of mass of the pixels that lie within 0.1
$R_{\sun}$ of the front of the flux rope and within $\pm 25^\circ$ of
the radial line that connects the sun center with the center of
mass. The velocity of the front, $v_{f}$ is calculated in the same
manner as $v_{CM}$ (\S~2.2). The comparison of the two velocity
profiles for some representative events is shown in
Figure~\ref{front}. Six of the eleven CMEs have profiles similar to
97/08/13 (self-similar expansion) or 97/10/30 (no expansion), while
five show a progressive flattening such as 97/04/13 or 97/11/01,
similar to that found in \citet{wood99}. Some theoretical flux rope
models also predict flattening of the flux rope as it propagates
outwards \citep{chen97, wood99}.
\section{Conclusions}

We have examined, for the first time, the energetics of 11 flux rope
CMEs as they progress through the outer corona into the heliosphere.
The kinetic and potential energies are computed directly from
calibrated LASCO C2 and C3 images, while the magnetic energy is based
on estimates from near-Earth in-situ measurements of magnetic
clouds. These results are expected to provide constraints on flux rope
models of CMEs and shed light on the mechanisms that drive such CMEs.
These measurements provide no information about the initial phases of
the CME (at radii below $\sim 2 R_{\odot}$). All the measurements and
conclusions hold for heights in the C2 and C3 fields of view; between
3 and $30 R_{\odot}$. The salient conclusions from an examination of
11 CMEs with a flux rope morphology are:

\begin{itemize}
\item
For relatively slow CMEs, which constitute the majority of events,
\begin{itemize}
\item
The potential energy is greater than the kinetic energy.
\item
The magnetic energy advected by the flux rope is
given up to the potential and kinetic energies, 
keeping the total energy roughly constant.
In this sense, these events are magnetically driven.
\end{itemize}
\item
For the relatively fast CMEs with velocities $\geq$ 600 km/s
(97/02/23, 98/06/02),
\begin{itemize}
\item
The kinetic energy exceeds the potential energy by the time they reach
the outer corona (above $\sim 15 R_{\sun}$).
\item
The magnetic energy carried by flux rope is significantly
below the potential and kinetic energies.
\end{itemize}
\end{itemize}

\acknowledgements We thank D. Spicer for the initial discussions that
led to this paper and the referee for his/her constructive comments. SOHO is an
international collaboration between NASA and ESA and is part of the
International Solar Terrestrial Physics Program. LASCO was constructed
by a consortium of institutions: the Naval Research Laboratory
(Washington, DC, USA), the University of Birmingham (Birmingham, UK),
the Max-Planck-Institut f\"ur Aeronomie (Katlenburg-Lindau, Germany)
and the Laboratoire d'Astronomie Spatiale (Marseille, France).

\newpage
\end{document}